\documentstyle[prl,aps]{revtex}
\tighten
\begin{document}
\draft

\title{
Classical Quantization of Local Hall Conductivity in 2D Ballistic Systems
       }

\author{ Yuli V. Nazarov and B. Mijling}
\address{
Delf University of Technology, Laboratory of Applied Physics and DIMES,
 2628 CJ Delft, the Netherlands
	}
\maketitle
\begin{abstract}
We study the linear Hall response of 2D ballistic system on inhomogeneous magnetic field.
We establish that in classical limit the Hall conductivity response on local magnetic field 
is quantized in units of $\alpha_H \equiv \frac{e^3}{2 \pi^2 \hbar c}$. 
The quantized value depends on the point where the field is applied, this dependence being irregular
in chaotic billiards. The phenomenon allows for direct tracing of special electron trajectories 
that belong to fractal repellor.
We discuss how quantum effects smooth the quantization.   
\end{abstract} 


Since the first ballistic constriction in 2D electron gas has been
formed \cite{First}
a great deal of experimental and theoretical attention has been devoted to
magnetotransport in such microstructures (see \cite{Carlo} for review). Several effects 
such as mesoscopic fluctuations \cite{Stone} can only be understood on 
quantummechanical basis. 
However the main features of magnetotransport
are to be comphehended in classical framework \cite{Carlo} and can be naturally  
explaned in language of classical electron trajectories.

An interesting new development  is to use the ballistic constriction as a sensitive 
detector of {\it inhomogenious} magnetic field.\cite{Geim} In this experimetal work, 
the constriction has been used to reveal the details of magnetization of a microscopic 
superconducting particle. However, the intrinsic physics of this detector also appears
to be interesting and complicated. There are several theoretical and numerical works 
tackling the problem of inhomogenous magnetic field. \cite{Peeters} In the present paper
we adopt the classical approach and concentrate on the Hall response
linear in inhomogeneous magnetic field applied.

We establish two main results: i. the linear Hall response
is quantized in units of $\alpha_H \equiv \frac{e^3}{2 \pi^2 \hbar c}$, the quantized value depending on the position in the
constriction ii. The lines that separate regions of different quantized values outline
the electron trajectories that belong to {\it fractal repellor}.\cite{Chaotic} 
The fractal repellor is a manyfold of electron trajectories orbiting inside the 
constriction for infinite time. This fact is of  importance since it opens the way for
direct observation of the chaotic nature of scattering in a ballistic constriction 
in a {\it transport} measurement. This is hard to believe since any transport quantity
results from averaging over extended regions of phase space. Thus one could expect that 
the chaotic dependence of the scattering result on initial conditions  is averaged out.

Let us start with the theory formulation for arbitrary multi-lead
ballistic constrictions. \cite{Butt} (Fig. 1). The currents $I_i$ in the
leads are related to the voltages by means of conductivity matrix:
$I_i=G_{ij} V_j$. It obeys Onzager relations $G_{ij}(H)=G_{ji}(-H)$ and
provides conservation of current: $\sum_i G_{ij}=0$. In classical limit,
the conductivity is determined from the master equation
\begin{equation}
\frac{\partial f}{\partial t} + \frac{{\bf p}}{m} \cdot
  \frac{\partial f}{\partial {\bf r}} -\frac{\partial U({\bf r})}{\partial {\bf r}}
\cdot \frac{\partial f}{\partial {\bf p}}
+ \frac{e}{m c} (
  {{\bf p}} \times {{\bf H}} ) \cdot \frac{\partial f}{\partial {\bf p}} = 0
\end{equation}
for distribution function $f({\bf r},{\bf p})$. We concentrate on
deviations of the conductivity matrix which are linear in $H$. Quite generally,
those can be presented if the form:
\begin{equation}
\delta G_{ij} = \frac{e^3}{2 \pi^2 \hbar^2 c} \int d^2 r N_{ij}({\bf r}) H({\bf r}).
\label{defe} 
\end{equation}
Now we are to demonstrate that $N_{ij}$ can only take {\it integer} values.
We expand distribution function in the form $f=f_0 +  f_1$, $f_1$ being proportional to the magnetic field, that gives equation
for $f_1$:
\begin{equation} 
\frac{\partial f_1}{\partial t} + \frac{{\bf p}}{m} \cdot
  \frac{\partial f_1}{\partial {\bf r}} -\frac{\partial U({\bf r})}{\partial {\bf r}}
\cdot \frac{\partial f_1}{\partial {\bf p}}= -\frac{e}{m c} (
  {{\bf p}} \times {{\bf z}})  \cdot \frac{\partial f_0}{\partial {\bf p}} H({\bf r}) 
\label{forf1}
\end{equation}
We can find the current generated by $f_1$ without explicitly solving
Eq. \ref{forf1}. To do this, we note that the right hand side of
(\ref{forf1}) can be presented as a sum over particle sources. We
introduce  the probability $P_i(x,y,{\bf p})$ to get to the lead $i$
while starting in point $x,y$ with momentum ${\bf p}$. Then the current
deviation can be presented in the form
\begin{equation}
\delta I_i = -\frac{e^2}{m c} \int \frac{d^2 p}{(2 \pi \hbar)^2} P_i({\bf r},{\bf p})({{\bf p}}
\times {{\bf z}})  
\cdot \frac{\partial f_0}{\partial {\bf p}} H({\bf r})
\end{equation}
Now we shall specify $f_0$. Since the system is ballistic, $f_0$ is
contributed by electrons coming from the leads. Since the probability
to come from $i-th$ lead is $P_i(x,y,-{\bf p})$, $f_0(x,y,{{\bf p}})=\sum_i P_i(x,y,-{\bf p}) f_F(\epsilon({\bf p}-eV_i)$. The deviation of the current 
is thus given by 
\begin{equation}
\delta I_i = -\frac{e^2}{m c} \sum_j
\int \frac{d^2 p}{(2 \pi \hbar)^2} P_i({\bf r},{\bf p}) ({{\bf p}}
\times {{\bf z}} ) 
\cdot \frac{\partial P_j({\bf r},-{\bf p})}{\partial {\bf p}}  f_F(\epsilon({\bf p}-eV_j))H({\bf r})
\label{integral}
\end{equation}

 We note that the electrons participating in transport are concentrated
in a narrow energy strip near Fermi level. That allows us to disregard the dependence
of $P$ upon energy or modulus of ${\bf p}$. $P$ becomes now a function of angle $\phi$ that 
determines the bearing of ${\bf p}$. Under this assumption, Eq.
\ref{integral} can be integrated over energy. We use the density of
states $\nu = m/\pi \hbar^2$.
This finally yeilds Eq. \ref{defe}
with 
\begin{equation}
N_{ij}=  \int_0^{2\pi} d\phi P_i(\phi) \frac{\partial P_j(\phi+\pi)}{\partial \phi} 
\label{winds}
\end{equation}
where the angle $\phi$ parametrizes direction of ${\bf p}$. Since the motion
of electrons is classical, $P(\phi)$ can only take two values: $0$ and $1$. So that
$N_{ij}$ is integer. We stress that this proof uses only the fact that
the scattering in the constriction is a classical one. It does not
depend on details of potential relief. Also, the minor modification of
the derivation shows that the result holds for an arbitrary electron
dispersion law provided the Fermi surface remains single-connected.

Does this mean that the linear Hall response is {\it physically}
quantized? No, not  necessarily. The physical, observable quantization
implies that $N(x,y)$ not only takes integer values at any point but also 
retains the same value in some finite neighborhood of this point. We will
see below that sometimes it is not the case.

Let us have a closer look at functions $P_i(\phi), P_j(\phi-\pi)$ (Fig. 1). 
We consider, without loss
of generality, only constrictions with smooth edges. 
Then  $P_i(\phi)$ develops a  typical self-similar chaotic stucture.\cite{Chaotic}
Although it can only take two values, $0$ and $1$, it changes  
infinite number of times between
these values. 
Moreover, in any finite neighborhood of each change one finds infinitely 
many other changes. 
In order to quantify the situation, one subdivides the whole interval of angle on 
smaller intervals of 
the length $2 \pi /K$, $K$ being big integer number. An interval can be 
"black" if $P_i=1$ for all its points, "white" if $P_i=O$ for all its points,
or "gray" if there is at least one change within the interval.
 With increasing resolution $K$, 
the total number of "gray" intervals $M$ scales as
\begin{equation}
M \simeq K^{d_H},
\end{equation}
$d_H$ being Hausdorf dimension of the fractal manyfold of changes. The dimension $d_H$ does not 
depend on a point where $P_i$ is measured. It is determined by the overall geometry 
of the constrictions.

Let us now analyze the expression (\ref{winds}) at finite resolution $K$.
First it seems that the answer is not determined at any finite $K$ since
the answer may depend on the behavior of $P$ within "gray" intervals. 
However, we notice that the expression (\ref{winds}) is completely determined provided
all "gray" intervals of $P_i$ are covered by "black" or "white" intervals
of $P_j$ and, similar, all "gray" intervals of $P_j$ are covered by "black" or 
"white" intervals of $P_i$. This stems from the structure of integrand: if $P_i$
is a constant within an  interval $(a,b)$, the contribution of this interval
equals $P_i(P_j(a)-P_j(b))$ and does not depend on the details of $P_j$.
Let us assume that for well-developed chaos positions of "gray" intervals for
$P_i$ and $P_j$ are statistically independent. We estimate the probability for 
all "gray" intervals be covered. Since the probability to find a "gray" interval
against a given interval is $M/K$, and the number of "gray" intervals is $M$, we obtain
\begin{equation}
P_{cover} = (1-\frac{M}{K})^M \approx \exp(-\frac{M^2}{K})
\label{area}
\end{equation}
in the limit of $M \rightarrow \infty$, $M/K \rightarrow 0$. Since $M \simeq K^{d_H}$,
we find two distinct regimes. If $d_H > 1/2$ $P_{cover}$ approaches $0$ 
with increasing resolution. This means that $N(x,y)$ can not be an integer in any
finite region of $x,y$ but rather shall exhibit continuous behavior.
In opposite case $d_H < 1/2$ $P_{cover}$ reaches $1$ so that $N(x,y)$ takes only integer
values that persists in  finite regions.

%chaotic scattering
Let us now reformulate the above statements stressing the relation between the changes of $P_{i,j}(x,y,\phi)$ 
and the trajectories
of electrons that exhibit {\it chaotic scattering} in the constriction. \cite{Chaotic}
The trajectory that starts at the point $(x,y)$ with the bearing $\phi$ 
at which 
a change occur may end neither in the lead $i$ ($P_i=1$)  
nor in any other lead ($P_i=0$).
Rather, it stays in the constriction for infinite
time. Let us now consider the trajectory of opposite bearing, $ \phi+\pi$. 
If for all infinite trajectories their counterparts of opposite bearing
are finite, we recover the above mentioned situation where all "grey"
intervals are covered by "white" or "black" intervals. Thus $N(x,y)$ is
defined and integer in this point.
Albeit this is not the case if in a given $(x,y)$ at some $\phi$
there is a trajectory that is infinite for both bearings, $\phi$ and
$\pi+\phi$,
so that $N(x,y)$ remains undefined in this point.
Owing to time reversal symmetry, this trajectory must correspond to a periodic
orbit in the constriction.
These periodic orbits form what is called a fractal repellor of the constriction.
Therefore the regions of constant $N(x,y)$
can be separated from each other only  by the trajectories of the fractal repellor. This is how these
trajectories can be seen. 

%Visulalisation
An imaginable visualization experiment would resemble those with STM or AFM. Let us bias
the constriction and measure its Hall response. As a source of magnetic field we can take a long
and sharp magnetic needle with one of its ends hanging close to the sample surface. We can scan the
whole constriction area with this needle measuring the Hall response $G_{ij}$ as a function of $(x,y)$
needle position. If we contrast the changes of the response we will see bunches of the fractal repellor
trajectories. (Figs. 2-4)
%Real simulations

To illustrate the visualization of the fractal repellor, we simulate classical electron trajectories
in the cross-like billiard with smooth edges (insert in Fig.2), these edges being quarter circles.
This system resembles the actual experimental setup. \cite{Geim} If the Hall conductivity
is defined in a conventional  way as the ratio of Hall current to the voltage drop across the
whole constriction, it is quantized in units of $\alpha_H/2$.
The chaotic  properties can be characterized by a single dimensionless parameter $\beta \equiv R/d$,
$R$ being circle radius, $d$ being the lead width. This parameter can be in principle changed 
{\it in situ} by means of gate electrodes. We stress that such a system corresponds to a classical model problem of chaotic scattering:
scattering by unpenetrable discs. Although a three-disk billiard received most of the attention \cite{Three}, 
the four-disk billiards have been thoroughly studied as well. \cite{Four} The Hausdorf dimension
denined as above was shown to change monotonically between $1/3$ ($\beta=0$) and $1$ ($\beta \rightarrow 1$)
with increasing $\beta$. The transition between discrete and continuous behavior of $N(x,y)$ takes place
at $\beta \approx 0.5$. We have chosen $\beta=1/3$ that corresponds to $d_H \approx 0.4$.

We have calculated approx. $100000$ periodic orbits, some of them being equivalent due to symmetry
of the problem. We plot them in Fig. 2, 3, 4 zooming on the constriction center with zoom factor
of $10$. We observe self-similar pattern which is characteristic for
chaos. Despite the huge number of the trajectories, and $d_H$ close to $1/2$, 
they do not uniformly cover the whole area of the constriction. Rather, 
they are grouped in bunches with well resolved space in between.  
We plot the local Hall conductivity along $y$ symmetry axis in right plane 
of each figure. That allows one to see the correspondence between
trajectories of the repellor and quantized values of $N(x,y)$.

Quantum mechanics is known to smooth peculiarities of the classical
chaos. \cite{Three} Although the representative quantum mechanical 
solution of the problem is hardly within the reach due to big number of
basis states required, we can draw simple estimations of quantum effects
in the billiards. We assume that the angle resolution $K/2\pi$ is 
limited by a typical diffraction angle $\phi_d = \sqrt{\lambda/L}$,
Here $\lambda$, $L$ are the electron wavelength and the typical size 
of the constriction, respectively. This implies that the regions of 
quantized values of $N(x,y)$ that would require better angular resolution,
remain unresolved and $N(x,y)$ is continuous there. Eq. \ref{area}
gives an estimation of relative area of these unresolved regions,
\begin{equation}
\Delta S/S \simeq \exp(-(L/\lambda)^{1/2-d_H}).
\end{equation}

In conclusion, we have predicted that the local Hall conductivity
of classical ballistic constriction may be quantized in values of 
$\alpha_H=\frac{e^3}{2 \pi^2 \hbar c}$. 
This may open up new ways to measure fundamental constants.
There is a transition between quantized and continuous behavior
of the local conductivity that takes place at $d_H=1/2$. The regions
of the same quantized $N(x,y)$ are separated by trajectories of
the fractal repellor, that provides the means for their visualization.

I am indebted to A. K. Geim for numerous discussions and his persistent 
interest in this work. Useful discussions with F. M. Peeters and
G. E. W. Bauer are greatly appreciated.
This work is a part of the research programme of the "Stichting voor
Fundamenteel Onderzoek der Materie"~(FOM), and I acknowledge the financial
support from the "Nederlandse Organisatie voor Wetenschappelijk Onderzoek"
~(NWO).

\begin{figure}
\caption{
Multi-lead ballistic constriction. The Hall response on magnetic field in the point $(x,y)$
is determined by probabilities $P_{i,j}(x,y, \phi)$. Their angular dependence is sketched on the right.
Grey intervals comprise fine chaotic structure of $P$ that can not be resolved at a given
angular resolution.
}
\label{fig1}
\end{figure}

\begin{figure}
\caption{
The fractal repellor trajectories in the constriction. Only a quarter of the constriction
is shown. The right panel presents quantized linear Hall response along the $y$ symmetry axis.
}
\label{fig2}
\end{figure}

\begin{figure}
\caption{
The same as in Fig 2. but at 10 times smaller scale.
}
\label{fig3}
\end{figure}

\begin{figure}
\caption{
The same as in Fig 2 at 100 times smaller scale.
}
\label{fig4}
\end{figure}
\end{document}